\documentstyle{mn}

\newif\ifAMStwofonts

\def\beq{\begin{equation}}
\def\eeq{\end{equation}}

\def\sc{\scriptstyle}
\def\Chi{{\raise 0.4ex\hbox{$\chi$}}}
\newcommand{\sub}[1]{_{\raise -0.3ex\hbox{$\sc #1$}}}

\ifoldfss
  \ifCUPmtlplainloaded \else
    \NewTextAlphabet{textbfit} {cmbxti10} {}
    \NewTextAlphabet{textbfss} {cmssbx10} {}
    \NewMathAlphabet{mathbfit} {cmbxti10} {} 
    \NewMathAlphabet{mathbfss} {cmssbx10} {} 
  \fi
  \ifAMStwofonts
    \ifCUPmtlplainloaded \else
      \NewSymbolFont{upmath} {eurm10}
      \NewSymbolFont{AMSa} {msam10}
      \NewMathSymbol{\upi}     {0}{upmath}{19}
      \NewMathSymbol{\umu}     {0}{upmath}{16}
      \NewMathSymbol{\upartial}{0}{upmath}{40}
      \NewMathSymbol{\leqslant}{3}{AMSa}{36}
      \NewMathSymbol{\geqslant}{3}{AMSa}{3E}

    \fi
  \fi
\fi 

\ifnfssone
  \newmathalphabet{\mathit}
  \addtoversion{normal}{\mathit}{cmr}{m}{it}
  \addtoversion{bold}{\mathit}{cmr}{bx}{it}
  \newmathalphabet{\mathbfit} 
  \addtoversion{normal}{\mathbfit}{cmr}{bx}{it}
  \addtoversion{bold}{\mathbfit}{cmr}{bx}{it}
  \newmathalphabet{\mathbfss} 
  \addtoversion{normal}{\mathbfss}{cmss}{bx}{n}
  \addtoversion{bold}{\mathbfss}{cmss}{bx}{n}
  \ifAMStwofonts
    \ifCUPmtlplainloaded \else
      \UseAMStwoboldmath
      \makeatletter
      \new@mathgroup\upmath@group
      \define@mathgroup\mv@normal\upmath@group{eur}{m}{n}
      \define@mathgroup\mv@bold\upmath@group{eur}{b}{n}
      \edef\UPM{\hexnumber\upmath@group}
      \new@mathgroup\amsa@group
      \define@mathgroup\mv@normal\amsa@group{msa}{m}{n}
      \define@mathgroup\mv@bold\amsa@group{msa}{m}{n}
      \edef\AMSa{\hexnumber\amsa@group}
      \makeatother
      \mathchardef\upi="0\UPM19
      \mathchardef\umu="0\UPM16
      \mathchardef\upartial="0\UPM40
      \mathchardef\leqslant="3\AMSa36
      \mathchardef\geqslant="3\AMSa3E
    \fi
  \fi
\fi 

\ifnfsstwo
  \DeclareMathAlphabet{\mathbfit}{OT1}{cmr}{bx}{it}
  \SetMathAlphabet\mathbfit{bold}{OT1}{cmr}{bx}{it}
  \DeclareMathAlphabet{\mathbfss}{OT1}{cmss}{bx}{n}
  \SetMathAlphabet\mathbfss{bold}{OT1}{cmss}{bx}{n}
  \ifAMStwofonts
    \ifCUPmtlplainloaded \else
      \DeclareSymbolFont{UPM}{U}{eur}{m}{n}
      \SetSymbolFont{UPM}{bold}{U}{eur}{b}{n}
      \DeclareSymbolFont{AMSa}{U}{msa}{m}{n}
      \DeclareMathSymbol{\upi}{0}{UPM}{"19}
      \DeclareMathSymbol{\umu}{0}{UPM}{"16}
      \DeclareMathSymbol{\upartial}{0}{UPM}{"40}
      \DeclareMathSymbol{\leqslant}{3}{AMSa}{"36}
      \DeclareMathSymbol{\geqslant}{3}{AMSa}{"3E}
    \fi
  \fi
\fi 

\ifCUPmtlplainloaded \else
  \ifAMStwofonts \else 
    \def\upi{\pi}
    \def\umu{\mu}
    \def\upartial{\partial}
  \fi
\fi

\def\LaTeX{L\kern-.36em\raise.3ex\hbox{a}\kern-.15em
    T\kern-.1667em\lower.7ex\hbox{E}\kern-.125emX}

\title{Galactic archaeology: IMF and depletion in the ``thin disk''}  

\author[K.-P. Schr\"oder and  B. Pagel]
{K.-P. Schr\"oder$^{1,2}$\thanks{E-mail: kps@star.cpes.sussex.ac.uk}, 
B. Pagel$^1$\\
$^1$Astronomy Centre, CPES, University of Sussex, Falmer, 
Brighton BN1 9QJ, UK \\
$^2$Technische Universit\"at Berlin, Zentrum f\"ur Astronomie und
  Astrophysik, PN~8-1, Hardenbergstr. 36, 10623 Berlin, Germany }

\date{Received Feb. 3, 2003; accepted ..., 2003}

\pagerange{\pageref{firstpage}--\pageref{lastpage}}
\pubyear{2003}

\begin{document}

\label{firstpage}

\maketitle

\begin{abstract}
We determine the initial mass function (IMF) of the ``thin disk'' 
by means of a direct comparison between synthetic stellar samples 
(for different matching choices of IMF, star formation rate SFR and 
depletion) and a complete (volume-limited) sample of single stars 
near the galactic plane ($|z| < 25$\,pc), selected from the 
Hipparcos catalogue ($d < 100$\,pc, M$_{\rm V} < +4.0$). 
Our synthetic samples are computed from first 
principles: stars are created with a random distribution of mass $M_*$ 
and age $t_*$ which follow a given (genuine) IMF and SFR($t_*$). They 
are then placed in the HR diagram by means of a grid of empirically 
well-tested evolution tracks. 
The quality of the match (synthetic versus observed sample) is 
assessed by means of star counts in specific regions in the HR diagram.
7 regions are located along the MS (main sequence, mass sensitive), 
while 4 regions represent different evolved (age-sensitive) stages 
of the stars. 

We find a bent slope of the IMF (using the Scalo notation, 
i.e., power law on a logarithmic mass scale), with 
$\Gamma_1 = -1.70 \pm 0.1$ (for $1.1 M_{\odot} < M_* < 1.6 M_{\odot}$) 
and $\Gamma_2 = -2.1 \pm 0.1$ 
(for $\approx 4 M_{\odot} > M_* > 1.6 M_{\odot}$).
In addition, comparison of the observed MS star counts with those of 
synthetic samples with a different prescription of the MS core 
overshooting reveals sensitively that the right overshoot onset is at 
$M_* = 1.50 M_{\odot}$. 

The counts of evolved stars, in particular, give valuable evidence of 
the history of the ``thin disk'' (apparent) star formation and
lift the ambiguities in models restricted to MS star counts.
A match of the evolved star counts yields a stellar depletion  
in the sample volume which increases with age (i.e., by apparent 
SFR(age)~/~SFR$_0$). A very good match is achieved with a simplistic 
diffusion approximation, with an age-independent diffusion time-scale
of $\tau_{\rm dif} = 6.3 \cdot 10^9$~yrs and a (local) 
SFR$_0 = 2.0 \pm 0.1$ stars (with $M_* > 0.9 M_{\odot}$) 
formed per 1000 years and (kpc)$^3$.  We also discuss this 
``thin disk'' depletion in terms of a geometrical 
dilution of the expanding stellar ``gas'', with $H_z(t_*) \propto 
\sigma_W(t_*)$. This model applies to all stars old enough to have
reached thermalization, i.e. for $t_* > 7 \cdot 10^8$~yrs and 
$H_z > 230$~pc. It yields a 
column-integrated (non-local) ``thin-disk'' SFR$_{\rm col}$ which 
has not changed much over time ($< 30$\%), 
SFR$_{\rm col} \approx 0.82$~*~pc$^{-2}$Gyr$^{-1}$ 
(i.e., stars with $M_* > 0.9 M_{\odot}$).
\end{abstract} 


\vspace{0.3cm}


\section{Introduction}

The case of the IMF has spurred a vast amount of literature over the
past 3 decades. We would like to mention only the classic work and 
reviews by Miller \& Scalo (1979) and Scalo (1986), together with 
Scalo (1998) and Kroupa (2001) for reference to the more recent work
in the field. The classical approach to derive the IMF of field stars
in the galactic disk follows a line of steps: (i)
derive the stellar distribution in absolute magnitude from the observed 
distribution in apparent magnitude, requiring the respective vertical 
space-density scale-height, and convert that into the luminosity 
function LF for MS stars, (ii) apply a theoretical mass-luminosity 
(m-L) relationship for obtaining the present-day mass function (PDMF),
and (iii) model the IMF on the PDMF, for a given SFR history. 
As discussed by Scalo (1998), each of these steps introduces 
specific uncertainties so that the case of the IMF is far from being
closed. Scalo (1998) and Kroupa (2001) even argue that there are 
systematic variations in the IMF, especially between the field star
IMF and the various IMF's derived from galactic stellar clusters. 

The availability of good parallaxes, as required for a precise field star 
LF, is mutually exclusive with the necessity of a large stellar sample. 
Therefore, early work on the LF had to rely on a statistical approach, 
based on the apparent magnitudes, or on photometric parallaxes (e.g., 
Gilmore \& Reid 1983). These approaches link the question of the IMF  
to that of the spatial stellar distribution, i.e., perpendicular to 
the galactic plane. Most notably, Gilmore \& Reid (1983) found that 
(i) the stellar spatial density scale heights increase with lower masses,
and that (ii) the global spatial distribution could be described by
a (denser) Galactic ``thin disk'' plus an extended Galactic ``thick disk''. 
Especially the first point is of importance when deriving a field star
IMF, as we will show below.

The great interest in the IMF is fuelled not only by the attempts 
to understand and model the star formation process(es). Finding the 
IMF is also closely linked to the question of galactic star formation 
history, i.e., whether and how the SFR has varied over time. Both IMF and 
SFR(t) are important ingredients of galaxy models since these quantities
determine such important aspects as the chemical evolution and the 
luminosity evolution of a galaxy (see, e.g., Dwek 1998, Gilmore 1999, 
Hensler 1999, Pagel 2001a,b). Furthermore, the IMF of more massive stars 
is of particular interest for the formation of star-burst regions
and stellar clusters, and for the evolution of young galaxies. Our Galaxy
would make an ideal test case for detailed galaxy evolution models, if
we could determine the IMF and SFR(t) with sufficient accuracy.
Already, we have quite a good account of the stellar population(s), 
the (local) kinematics and chemical composition as a function 
of age.

A general problem in this context is the inherent ambiguity of any
interpretation that is restricted to MS stars: Towards lower stellar
masses, the mean age of a MS star increases accordingly. Consequently, 
assuming a different time-dependence of the rate of star formation  
(i.e., increasing, or decreasing with age) has the same effect on 
the expected numbers of stars on the lower MS as changing the IMF
accordingly. This degeneracy of SFR(t) and lower IMF (see, e.g., Binney 
et al. 2000) allows a variety of false combinations of IMF and SFR(t) 
to match the PDMF of MS stars. It is therefore important to 
lift this degeneracy by extracting additional, specifically
age-sensitive information from the HR diagram which is buried 
in the properties and numbers of the evolved stars. 

Matters are further complicated by the galactic disk dynamics. Relevant 
properties of the star formation process are the genuine IMF and SFR
at the time of the star formation and in a defined volume on the 
galactic plane  -- i.e., within the vertical extent of the star 
formation sites. However, present-day counts of stars in a 
solar neighbourhood volume translate into an {\it apparent} 
SFR which decreases towards larger age (Schr\"oder \& Sedlmayr 
2001). This depletion of stars in the 
galactic plane is caused by a dilution of the ``heated stellar gas'' 
which is expanding vertically (see below). If 
not accounted for, this effect would make a spurious, systematic 
difference to, e.g., the steepness of any IMF derived from a 
volume-limited stellar sample. Consequently, the above-mentioned 
classical studies of IMF and SFR have used the
invariant column-integrals. Their dependence on stellar mass and age is 
equivalent to that of the genuine IMF and the depletion-corrected
SFR in the volume. This column-integral approach, however, has the 
disadvantage of being no more accurate than the required and presently
quite limited knowledge of space-density scale-heights as a function 
of stellar age.

For this reason, we use the direct approach to star formation, 
based on the stellar records in a well-defined (local) volume, 
and an interpretation in terms of a matching synthetic sample. 
Synthetic stars are randomly placed in a well-tested evolution grid, 
and their numbers are (statistically) 
defined by the given (genuine) IMF and (apparent) SFR$_{\rm ap}$ 
in the volume. In previous studies we have shown that 
counts of specific synthetic stars compare well with the respective, 
observed present-day star counts (Schr\"oder 1998; Schr\"oder \& 
Sedlmayr 2001). The potential of this approach lies in its 
simplicity. We do not need to determine a LF, nor to use a (m-L) 
relation -- this is automatically accounted for by 
our synthetic sample creation, as is the mass-dependent fraction of 
post-AGB stars which are left out by a MS-based PDMF.  
The genuine SFR(t) in the volume is then derived from a depletion 
description (see below), for which a maximum of age-sensitive 
information is derived from the stellar sample. 

Studies based on a similar approach have been undertaken by 
Bertelli \& Nasi (2001), who, in addition, account for a variety of 
metallicity (see also Bertelli et al. 1992), and by Binney et al. (2000). 
However, these authors did not exploit the specific,
age-sensitive information provided by the counts of evolved stars.
In fact, the models of Bertelli \& Nasi (2001) show an excess of the 
number of evolved stars by (relative to the MS stars of same initial mass) 
a factor of 1.5, while the problem of stellar diffusion away from
the galactic plane has not been addressed. 

Consequently, the work presented here pays special attention to three
points which appear crucial for an improvement over previous
work: (1) In section 2, we describe a large, volume-limited sample 
of stars from the Hipparcos catalogue, which we selected for good 
proximity to the galactic plane ($|z| < 25$~pc) in order to avoid 
any selective under-representation of young, massive stars at larger 
$|z|$. (2) In section 3, we study the stellar depletion in the galactic 
plane. A simplistic but well-matching diffusion approximation with an 
age-independent diffusion time is compared with a more physical 
model. The latter describes the depletion in terms of geometrical 
dilution caused by the stellar kinematics: stars old enough 
($t_* > 7 \cdot 10^8$yrs) to have ``thermalized'' are 
expanding to larger scale heights exactly as their vertical
velocity distribution increases with age. We find a semi-empirical
prescription of the stellar space-density scale-heights in agreement
with the available observational evidence.
(3) In section 4, we describe the computation of our synthetic samples 
and how we assess a match with the observed stellar sample. Special
consideration is given to the counts of evolved stars, in order to 
constrain the right choice of IMF, SFR(t) and depletion as well as
possible.


\section{Complete samples of single stars on and near the galactic plane}

\begin{figure*}
\vspace{12cm}
\includegraphics{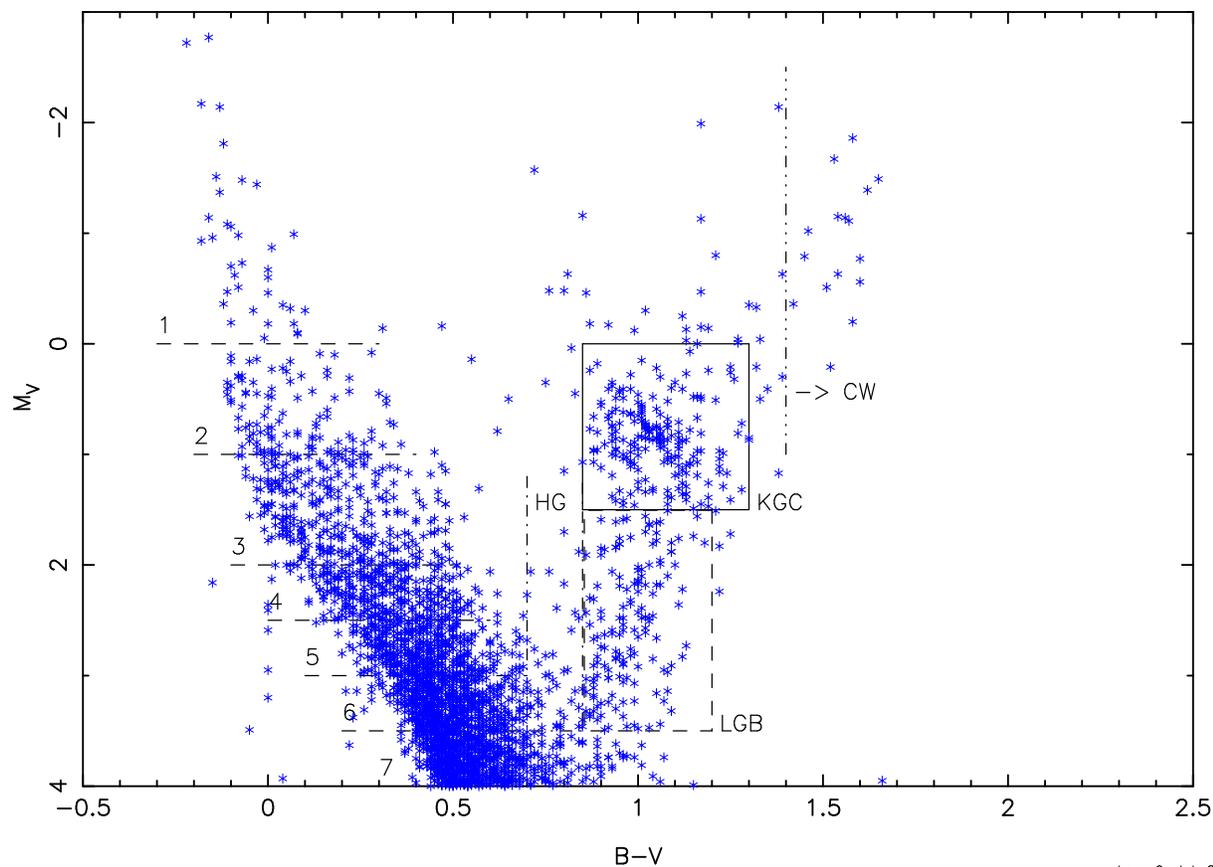}
\caption{The observed HR diagram of all single stars which are within
25\,pc to the galactic plane, less than 100\,pc away, and brighter than
M$_{\rm V} = 4.0$, as obtained from the Hipparcos catalogue entries. 
Also shown are the regions used for characteristic star counts 
(see text), 7 along the main sequence and 4 for evolved stages. }
\end{figure*}

In principle, we use the same volume-limited (i.e., $d < 100$~pc)
sample of single stars, based on the Hipparcos catalogue, 
as for earlier work (Schr\"oder 1998), but solving the problem with
consistency: In comparison to the very local sample with
$d < 50$~pc (used in Schr\"oder \& Sedmayr, 2001), the 100pc-volume provides
a comfortably larger statistical stellar basis, but it also comprises 
larger distances $z$ from the galactic plane which already make a difference 
to the PDMF of massive stars. Therefore, this investigation now discriminates 
against all stars with $|z|>25pc$ within 100~pc, which still leaves us
with larger, statistically more significant star counts than from 
the 50pc-sample. Star counts from neighbouring space are used, where
possible, to derive space-density scale-heights for specific groups of 
stars (see Table 1).
 
For computing the $z$ values, we adopted an offset of the Sun of 15pc 
north of the galactic plane (Cohen 1995), which intersects the spherical 
sample volume around the Sun ($R = 100$pc, $V = 4.19 \cdot 10^6$pc$^3$) 
off-centre. In order to keep the star counts of the neighbouring space 
reasonably large, we devide the spherical volume into slabs of (A) 
all stars within $|z| < 25$pc ($V_A = 1.50 \cdot 10^6$pc$^3$), (B) all stars
between 25pc$ < |z| < 50$pc ($V_B = 1.31 \cdot 10^6$pc$^3$), and (C) 
those stars in the polar sections of the 100pc sphere, 
up to $|z| < 85$pc and 115pc, respectively ($V_C = 1.38 \cdot 10^6$pc$^3$).

\begin{table*}             
\caption[]{Characteristic star counts in the space slabs A, B and C, 
for different regions in the HR diagram along the main sequence (MS) 
and for different groups of evolved stars (see Fig. 1 and text). The 
bottom line gives the approximate scale heights of the respective
stellar number densities.}

\begin{tabular}{llrrrrrrrcrrrr}
\hline 
Sample & $z$ [pc] & MS1 & MS2 & MS3 & MS4 & MS5 & MS6 & MS7 
                  &  TN &  HG & KGC & LGB & CW \\
\hline 
A & [0--25] & 38  & 106 & 317 & 327 & 484 & 712 & 850 
                  & 3348 & $<49$ & 205 & 157 & 16 \\
B &[25--50] & 13  &  79 & 227 & 261 & 365 & 581 & 743 
                  & 2736 & $<53$ & 171 & 146 & 21 \\
C &[50--115]& 20  &  63 & 212 & 260 & 374 & 644 & 835
                  & 2921 & $<69$ & 175 & 156 & 22 \\  
\hline
 &$H_{z,*}\approx$& 60: & 135 & 160 & 320: &270: & $>400$ & --- 
                  & $>400$ & --- & $>400$ & --- & --- \\          
\hline
\end{tabular}  \end{table*}       

Stellar population densities in the HR diagram are assessed by 
means of star counts in 7 regions along the main sequence (MS1 to MS7,
sensitive to the initial mass function IMF), and in 4 regions representative 
of different evolutionary stages and stellar ages, containing information 
on the star formation history SFR($t_*$) and the diffusion losses over
time: the Herzsprung gap (HG), the K giant 
clump (KGC), the lower giant branch (LGB) and stars with cool winds (CW), 
with border-lines as shown in Fig. 1. The observed HG counts are particularly 
affected by any unrecognized binaries with a composite colour and must be 
considered as upper limits, only.

Since the Hipparcos catalogue is complete at least for stars 
brighter than V\,=\,7.3 (in some regions of the sky even
down to V\,=\,9, see Perryman et al. 1997), completeness of 
volume-limited samples is automatically given for all stars within 
50pc which are brighter than M$_{\rm V}=4$.
Towards 100pc distance, however, star counts of regions MS6 and MS7 
(with M$_{\rm V}=3.0$ to 4.0) would be somewhat incomplete. By applying 
the (unaffected) star count ratios in the 50pc sample (as from Schr\"oder \& 
Sedlmayr 2001), especially to MS6, MS7 and the total number of stars (TN)
as a multiple of MS5, we can make a good estimate of the missing fractions. 
For the stars within $|z|<25$pc of the galactic plane, we find 
corrections for the counts of MS6 by 4\%, of MS7 by 25\%, and of TN by 8\%.
These figures need to be added for a comparison with the synthetic samples 
discussed in section 4.

Normalizing the characteristic star counts to their respective volumes
(A, B and C as given above), we derive approximate, local scale heights,  
as listed in the bottom line of Table 1.  In order to handle any intrinsic 
fluctuations of the star counts, we applied a comparison 
of the ratios $N_{*\rm,A}/N_{*\rm,B}$ and $N_{*\rm,A}/N_{*\rm,C}$ with 
volume integrals over exponential density distributions for different 
scale heights. For the sake of a more realistic description, we assume 
a constant space-density around the galactic plane
for $|z| < 10$~pc.

The small space-density scale-heights of the stars in groups MS1 to MS3  
confirm that the young and more massive 
stars are much more concentrated to the galactic plane, and that their 
correct assessment requires a volume restriction in $z$. 
For MS6, MS7 and LGB, in fact, stellar number 
densities increase slightly from B to C. This local deficiency of stars 
appears to correspond with the low-$z$-structure found by Kuiken \& Gilmore
(1989). Hence, the $H_z$-values stated in Table 1 should be larger than 
what is typical for the galactic thin disk over a larger $z$-range.  
Clearly, a complete sample covering a larger 
volume, as will be available from the future GAIA 
(e.g., Perryman 2002) and DIVA (Bastian et 
al. 1996, Seifert et al. 1998) missions, is highly desirable.


\section{Vertical space density structure and ``thin disk'' dynamics}

A lot of observational work has been done on the {\it stellar 
space-density scale-heights} $H_z$ vertical to the galactic plane. 
Galactic structure studies on a wide range of stars have come up with 
the concept of a ``thin'' disk and an old ``thick'' disk, with 
scale heights of $\approx$~250~pc to 400~pc (thin disk, e.g., 
Gilmore \& Reid 1983, Kent et al. 1991, Ojha et al. 1996,
Siebert et al. 2002) versus 700~pc to 800~pc (thick disk, 
e.g., Ojha et al. 1999, Robin et al. 1996, Kerber et al. 2001, 
Phleps et al. 2000). However, a single scale height
does not describe the ``thin disk'' stellar distribution adequately. 
Rather, each particular class of stars has a different scale height 
which increases with the respective mean stellar age. Evidence 
comes from selective star counts which focus on narrow age ranges, 
though specific values may still be quite uncertain:

(i) From an interstellar extinction study, Vergely et al. (1998) find
a scale-height of the local ISM density and clouds, the regions of star 
formation, to be between 35~pc and 55~pc. 

(ii) For young and massive stars, scale-heights between 
25~pc and 65~pc have been found (Reed 2000, for OB stars), as well as
45~pc (Conti \& Vacca 1990, WR stars). 

(iii) From a study of all F-type stars 
within a distance of 50~pc, which (on average) are half-way through their
main-sequence life, Marsakov \& Shelevev (1995) find a scale-height of 
160~pc. 

(iv) For thin disk K dwarfs, Kuiken \& Gilmore (1989) find a scale-height of
249~pc for $z>300$~pc. Nearer to the galactic plane, their density data show 
pronounced low-$z$-structure. A rough average would yield a 
significantly larger scale-height.

(v) For the galactic disk subgiants, Mendez \& van Altena (1996) find 
a red-giant scale-height of 250~pc. Many of these are He-burning 
K-giant-clump stars, represented by our group KGC. 

(vi) Jura \& Kleinmann (1992a,~b) studied K and M-type (O-rich) variable 
AGB stars and Miras. For periods of $P=300...400$ days, indicative of
a larger luminosity and, hence, less small mass (around 1.1, possibly 1.2
$M_{\odot}$) and lesser age, they find a typical scale-height of 250~pc. 

(vii) For the supposedly less luminous, and therefore presumably slightly older
group of stars with $P=200...300$ days, the same authors estimate 500~pc.
In addition, Jura (1994) finds a scale-height of 600~pc for the O-rich, 
short-period Miras, which he estimates at a mass as small as about 
$1.0 M_{\odot}$.
 
(viii) for RR Lyrae stars with a moderate metal under-abundance ([Fe/H]$>$-1)
Layden \& Andrew (1995) find a scale height of 700~pc. As expected from
their abundance, this places them at the transition from thin to thick disk.
 
A similar range of scale-heights emerges from a study of LMC foreground 
stars, grouped by different spectral class and metallicity, by 
Oestreicher \& Schmidt-Kaler (1995). These authors find 105~pc for 
early-type stars, to over 500~pc for the metal-poor, late-type giants.

In line with these observations, an analytical description of the 
stellar dynamics suggests age-dependent scale heights which increase
in proportion with the vertical stellar velocity distribution $\sigma_w$:

After a few oscillations at right angles to the Galactic plane, stars
reach a state of dynamical equilibrium governed by the potential gradient
at right angles to the plane and by the velocity dispersion (Oort 1932).
With the progress of
time, the velocity dispersion increases from the effect of encounters with
interstellar clouds; alternatively, the oldest stars, belonging to the
Thick Disk,  are born with a high velocity dispersion, which they still
retain, either as a result of being born from  collapsing gas clouds that 
had not yet undergone complete dissipation of their potential energy
(Eggen, Lynden-Bell \& Sandage 1962; Burkert, Truran \& Hensler 1992), 
or because of merger events in the early history of the disk 
(Robin et al. 1996; Pilyugin 1996). 
Either way, the velocity dispersion increases with age 
(Wielen 1977; Lacey 1984; Sommer-Larsen \& Antonuccio-Delogu 1993)
resulting in an increase in the width of the density distribution
as a function of $z$.

By Jeans's theorem, the distribution function in phase space in a steady
state is a function of isolating integrals of the motion (Jeans 1915,
Lynden-Bell 1962).
In the approximation where $z$-motions are considered as uncoupled from 
motion in the plane, the only isolating integral is the vertical energy

\begin{eqnarray}
E = \frac{1}{2}w^2 + \phi
\end{eqnarray}

and the distribution function (assumed to be gaussian for an isothermal
stellar population) is

\begin{eqnarray}
f(z, w) = \nu(z)\frac{1}{\sigma_w\surd(2\pi)}e^{-(w^2 +2\phi)/2\sigma^2_w}
\end{eqnarray}

where $\nu$ is the volume density at height $z$, $w$ the velocity
along the $z$-axis and $\sigma_w$ the corresponding velocity dispersion.
Consequently the total volume density is

\begin{eqnarray}
\nu(z)= \int_{-\infty}^{\infty}f(z, w) dw = \nu(0) e^{-\phi/\sigma^2_w}.
\end{eqnarray}

Kuijken \& Gilmore (1989) have given an expression for $\phi(z)$:

\begin{eqnarray}
\phi = 2 \pi G [\Sigma_0(\sqrt{z^2+D^2}-D)
+ \rho_{\rm eff} z^2] ,
\end{eqnarray}

where $\Sigma_ 0$ is the mass surface density of the disk, $D$ a 
length scale defining the transition from the linear regime 
$K_z \propto z$ to $K_z = $ const. and $\rho_{\rm eff}$ is the 
contribution of the dark halo to the
local density of gravitating matter. This expression
can be expanded (for $z^2<D^2$) as a binomial series

\begin{eqnarray}
\phi & \simeq & 2\pi G\left[\Sigma_0 
D\left(\frac{1}{2}\frac{z^2}{D^2}-\frac{1}{8}
\frac{z^4}{D^4}+ \ldots\right)+\rho_{\rm eff}z^2\right]  \nonumber   \\
 & = & 2\pi G\left[\left(\frac{\Sigma_0}{2D} + \rho_{\rm eff}\right)z^2
- \frac{1}{8} \Sigma_0 D(\frac{z}{D})^4 + \ldots\right].
\end{eqnarray}

The first term, $\phi\propto z^2$, corresponds to simple harmonic motion
where from Eq 3,

\begin{eqnarray}
\frac{\nu(z)}{\nu(0)} = e^{-A^2 z^2 / \sigma^2_w},
\end{eqnarray}

where

\begin{eqnarray}
A^2 \equiv  2 \pi G\left(\frac{\Sigma_0}{2D} + \rho_{\rm eff}\right)
\end{eqnarray}

and in this approximation  the column density of a tracer population by
number is related to the
central density by      

\begin{eqnarray}
N = \int_{-\infty}^{\infty} \nu(z)dz =\nu(0)\frac{\sigma_w}{A}\surd\pi,
\end{eqnarray}

or

\begin{eqnarray}
\nu(0) = \frac{A}{\sigma_w\surd\pi}N,
\end{eqnarray}

i.e. we have a gaussian distribution with an effective
scale height proportional to $\sigma_w$; in a higher approximation to
$\phi(z)$, Eq (3) must be integrated over $z$ numerically.

The values of the parameters $\Sigma_0$, $D$ and $\rho_{\rm eff}$ have
been discussed by Kuijken \& Gilmore (1989) and very recently by 
Siebert, Bienaym\'e and Soubiran (2002), among others. 
$\Sigma_0/D$ is rather well determined
at 0.18 $M_{\odot}$ pc$^{-3}$ and $\rho_{\rm eff}\simeq 0.02$ is a
relatively small correction, but there is a degeneracy in the determination
of $\Sigma_0$ and $D$. Siebert et al. favour $D \approx 400$ pc, by contrast to
Kuijken \& Gilmore's value of only 180 pc. The difference is crucial to the
range of validity of Eqs 6, 8 and 9. Taking $D=500$ pc, the quadratic
approximation to $\phi$ is accurate to better than 10 per cent up to
$z=350$ pc, which would make it applicable to virtually all thin-disk
stars in our neighbourhood, whereas with $D=180$ pc, the situation would 
be very different. The actual scale heights of the more numerous, 
less massive stars presented above, however, suggest that the approximation
is indeed valid for our sample.
                                          
In the approximation of the density distribution $\nu(z)$ being an
exponential function of $z$ with a scale height $H_z$, we have the 
simple relation $N = 2 \cdot H_z \cdot \nu(0)$. In combination with Eq (9) 
it then follows that $H_z(t_*)$ expands $\propto \sigma_w(t_*)$ and 
$\nu(0)$ dilutes $\propto 1/\sigma_w(t_*)$ with age and
increasing vertical velocity dispersion. 

For $\sigma_w(t_*)$ as a 
function of stellar age, we take the simple (constant-diffusion) solution 
of Wielen (1977), $\sigma_w = (a_W + b_W \cdot t_*)^{0.5}$.
For $t_* > 7 \cdot 10^8$~yrs, we find that Wielen's empirical data 
points, as well as those of Sommer-Larsen \& Antonuccio-Delogu (1993), 
can be well enough represented by $a_W = 72$~(km/s)$^2$, $b_W = 
72$~(km/s)$^2/10^9$yrs.  

Finally, with $H_z \propto \sigma_w(t_*)$ and in consideration of
the above listed empirical stellar space density scale heights, we 
have adopted the following prescription (in pc)
as a function of age (in yrs):

\begin{eqnarray}
H_{z,*} = 177 \cdot (1.0 + t_*/10^9)^{0.5}, ~~t* > 7 \cdot 10^8
\end{eqnarray}

In particular, scale heights range from around 265\,pc (about the 
average half-thickness of the thin disk according to Ojha et al. 1996,
reached around an age of $t_* \approx 1.2 \cdot 10^9$~yrs) to about
560\,pc for the very oldest ($9 \cdot 10^9$ yrs) thin-disk stars. 

The present-day, young stars are formed very near the galactic plane, 
with the scale-height of the ISM clouds ($H_{z,0} \approx 45$~pc, see 
above), and drift away with increasing age. As long as their kinematics 
are not thermalized (for $t_* < 7\cdot 10^8$, $H_z < 230$\,pc), we 
prescribe their scale-heights mainly by a quadratic superposition of the 
initial scale-height $H_{z,0}$ and a term which is linear in age (in years), 
as:

\begin{eqnarray}
H_{z,*} = (45^2 + (t_* / 3.105 \cdot 10^6 {\rm yrs})^2)^{0.5}, 
~~t_* < 7 \cdot 10^8  
\end{eqnarray}

The resulting values are in reasonable agreement with the scale-heights 
obtained directly from the corresponding age-groups of Hipparcos stars
(i.e., MS1, MS2, and MS3 in Table 1).


\section{``Thin disk'' depletion by dilution: expansion of the stellar ``gas''}

Near the galactic plane, the vertical gravitational net force is small
enough to be neglected. In an earlier study (Schr\"oder \& Sedlmayr 2001), 
we therefore suggested a
very simplistic model in which depletion was prescribed as a diffusion
effect, with a single diffusion-time scale $\tau_{\rm diff}$ for all stars
(now referred to as model a). A value of $6.3 \cdot 10^9$yrs, which we 
here confirm to give a good match of the observed star counts,  
leads to a 50\% depletion of stars aged $4 \cdot 10^9$yrs. To 
give another example of the significance of this depletion: the ratio of
less massive giant stars (aged around $5 \cdot 10^9$yrs) over 
their MS counterparts (of average age $2.4 \cdot 10^9$~yrs) 
is reduced by a factor of 1.5 (which is of the order of the mismatch 
found with synthetic samples by Bertelli \& Nasi, 2001).

For the IMF, we use the conventional power-law notation (Scalo, 1986) 
that counts on a logarithmic mass scale: 

\begin{eqnarray}
{\rm IMF}(M_*) = dn_*/d\log{M_*} \propto M_*^{\Gamma}  
\end{eqnarray}

\noindent To allow for a bent slope, we use a $\Gamma_1$ for
masses smaller than $1.6 M_{\odot}$, and a $\Gamma_2$ for larger masses.
The mass-range of this study is limited by the observed sample to
about(effectively) $4 M_{\odot}$ at the upper, and to about $1.1 M_{\odot}$
at the lower end. 

Our principal star formation observable is the {\it apparent} 
SFR$_{\rm ap}(t_*)$ in the sample volume. As mentioned before, it appears
to have been much lower in the past than at present (SFR$_0$) due 
to a dilution effect of the expanding stellar ``gas''. Therefore, we 
introduce a depletion 
factor of $F_{\rm depl}(t_*) =$ SFR$_{\rm ap}(t_*)$/SFR($t_*$) relative to 
the genuine star formation rate SFR($t_*$) in the volume, at the time of 
the star formation. In addition, we consider possible slow changes in the 
genuine SFR by an exponential factor, relative to the present-day star 
formation rate SFR$_0$, defined on the reversed time-scale of the 
stellar age $t_*$ (in years):

\begin{eqnarray}
{\rm SFR}_{\rm ap}(t_*) = F_{\rm depl}(t_*) \cdot {\rm SFR}_0 
                          \cdot e^{\gamma t_*/10^9}
\end{eqnarray}

\noindent Our values of the SFR (in the volume) are given in stars 
formed per year which have a mass larger than $0.9 M_{\odot}$, 
in the local sample volume A (see previous section) of 
$1.5 \cdot 10^6$pc$^3$. For our simplistic diffusion model (a), 
the depletion factor $F_{\rm depl}$ (for any given number $n_*$ of 
similar stars of a specific age and mass) is itself 
a $t_*$-dependent, exponential term 
(in Schr\"oder \& Sedlmayr 2001, both these terms were simply 
merged into one).

A more physical description of the depletion is given in terms 
of a geometrical dilution of stellar space density into the column
(model b). According to Eq 9, the depletion factor is then proportional 
to the vertical velocity distribution or, equally, to the scale-height 
of a group of stars of a given age:

\begin{eqnarray}
F_{\rm depl}(t_*) = H_{\rm ref} / H_z(t_*)
\end{eqnarray}

This model (b) applies, once (1) the stellar ``gas'' has thermalized (i.e., if
$t_* > 5 \cdot 10^8$yrs or more), and (2) if the depletion by dilution into 
the column is not compensated for by any fill-in from radial mixing in the 
disk. However, for the past $7 \cdot 10^8$yrs, significant fill-in is
evident: From this more recent time-span we would expect a 
dilution by a factor of about 5 (from the scale-height of the 
star-formation layer, about 45~pc, to $H_z(t_*=7 \cdot 10^8)
= 230$~pc. Instead, the respective $F_{\rm depl}(t_*)$ required by a 
matching synthetic sample (see below) is about 0.85 or even closer to 1. 

To match this depletion at the onset of geometrical dilution 
($7 \cdot 10^8$yrs), $H_{\rm ref}$ must be 195~pc.
For all younger ($t_* < 7 \cdot 10^8$yrs) 
stars we use the simple exponential approximation 
of the depletion factor this time with 
$\tau_{\rm dif} = 4.5 \cdot 10^9$yrs), guided by a smooth
transition with the depletion derived from dilution.

For a quite limited volume around the Sun, a significant fill-in over the 
more recent past must be expected: At present, it is located right in between 
two spiral arms. Consequently, the present-day, local SFR is much below the
global, average SFR for the ``thin disk''. As we look towards older stars, 
radial mixing changes the observed SFR from a local (for the 
youngest stars) to a non-local (after 2 to 3 orbital periods) 
quantity.  

A quantity more relevant for the physics of the galactic disk is the
column-integrated star formation rate, which relates to the apparent
local SFR in the volume, our principal observable, by:

\begin{eqnarray}
{\rm SFR}_{\rm col}(t_*) \approx 2 \cdot H_z(t_*) \cdot {\rm SFR}_{\rm ap}
\end{eqnarray}

This column-integrated SFR is then related to the {\it real} local star 
formation in the volume (allowing here for a possible exponential 
dependence on age) and the subsequent depletion, simply by:
 
\begin{eqnarray}
{\rm SFR}_{\rm col}(t_*) \approx 2 \cdot H_z(t_*) \cdot 
               F_{\rm depl}(t_*) \cdot {\rm SFR}_0  
                                 \cdot e^{\gamma t_*/10^9}
\end{eqnarray}

The SFR$_{\rm col}(t_*)$ derived from
our model first increases with look-back time (driven by the above-mentioned
fill-in from radial mixing which causes scale-heights of the younger stars
to increase steeply with age), until from an age of $7 \cdot 10^8$yrs it 
then settles on a disk-averaged value (see below). This is about 4 times 
larger than that of the low present-day, 
interarm star formation. The same effect would be observed, if the 
effective star formation layer had just been more expanded in the past
(as noted by Binney et al. 2000).
   

\section{Synthetic stellar samples and the IMF in the galactic plane}

\begin{table*}             
\caption[]{Characteristic star counts $N_i$ of different synthetic
samples and their deviations from observed values, in units of the
statistical variation $\sqrt{n_{i,\rm obs}}$, plus the mean deviation
$|\delta|_{\rm av}$ (double weight is given to KGC and LGB counts). 
Under the best-matching, simple diffusion model, synthetic samples based
on different evolution grids and thin disk ages are listed, followed 
by those derived with depletion by (mainly) dilution. A constant SFR
($\gamma = 0$) gives a better match than SFRs decreasing from an initially 
30\% ($\gamma = 0.03$) and 70\% ($\gamma =0.06$) higher value. }

\begin{tabular}{r|cccccccr}
\hline 
Model &{\bf MS1}&{\bf MS2}&{\bf MS3}&{\bf MS4}&{\bf MS5}&{\bf MS6}&{\bf MS7}&\\
      &   &  &  &{\bf HG} &{\bf KGC}&{\bf LGB}&{\bf CW} & $|\delta|_{\rm av}$\\
\hline 
{\bf Obs.}&{\bf 38}&{\bf 106}&{\bf 317}&{\bf 327}&{\bf 484}&{\bf 740}
&{\bf 1068}& \\ 
      &   &  &  & ($<49$) &{\bf 205}  & {\bf 157} & {\bf 16} & --- \\
\hline
$\tau_{\rm dif} = 6.29$ & $37^{(-0.10)}$ & $109^{(+0.29)}$ & $315^{(-0.14)}$ 
& $325^{(-0.13)}$ & $495^{(+0.48)}$  & $739^{(-0.04)}$ &$1077^{(+0.27)}$ & \\
 & & & & 44 &$209^{(+0.28)}$ & $149^{(-0.61)}$ &$18^{(+0.40)}$ &{\bf 0.30}\\
\hline

ovo=1.8 &$40^{(+0.25)}$ & $108^{(+0.17)}$ & $317^{(+0.0)}$&$260^{\bf(-3.7)}$ 
& $472^{(-0.50)}$ & $770^{\bf(+1.1)}$ &$1076^{(+0.23)}$ & \\
 & & & & 44 & $197^{(-0.59)}$ &$179^{\bf(+1.8)}$& $20^{(+1.0)}$ &{\bf 0.98}\\

ovo=1.6 & $40^{(+0.25)}$  & $109^{(+0.27)}$ & $325^{(+0.44)}$&$311^{(-0.87)}$ 
& $451^{\bf(-1.5)}$&$764^{(+0.87)}$ &$1062^{(-0.17)}$ &  \\
 & & & & 42 &$206^{(+0.11)}$  & $162^{(+0.36)}$ &$19^{(+0.65)}$ &{\bf 0.50}\\

ovo=1.4 &$37^{(-0.13)}$ & $112^{(+0.58)}$ & $322^{(+0.29)}$ & $318^{(-0.50)}$ 
& $539^{\bf(+2.5)}$  & $751^{(+0.41)}$ &$1064^{(-0.13)}$ &  \\
 & & & & 39 &$200^{(-0.36)}$ & $147^{(-0.77)}$ &$22^{(+1.5)}$ &{\bf 0.69}\\

$t_{\rm 9max}10$  & $38^{(-0.06)}$&$107^{(+0.14)}$&$311^{(-0.33)}$
  &$330^{(+0.19)}$&$507^{(+1.05)}$&$716^{(-0.89)}$ &$1067^{(-0.04)}$ & \\
  & & & & 39 &$199^{(-0.39)}$ &$159^{(+0.18)}$&$18^{(+0.44)}$ &{\bf 0.36}\\

$t_{\rm 9max}8$  & $41^{(+0.45)}$&$112^{(+0.58)}$&$317^{(-0.02)}$
  &$324^{(-0.17)}$&$498^{(+0.65)}$&$730^{(-0.36)}$ &$1079^{(+0.34)}$ & \\
  & & & & 48 &$191^{(-0.98)}$ &$142^{(-1.20)}$&$16^{(+0.10)}$ &{\bf 0.59}\\


$\tau_{\rm dif}\infty$, $\Gamma_1-0.5$& 
                      $42^{(+0.68)}$ & $108^{(+0.16)}$ & $311^{(-0.34)}$ 
& $329^{(+0.10)}$ &$519^{\bf(+1.59)}$ &$744^{(+0.15)}$ &$1073^{(+0.16)}$ & \\
& & & &56&$243^{\bf(+2.6)}$ & $220^{\bf(+5.1)}$ &$31^{\bf(+3.7)}$&{\bf 1.9}\\
\hline

Dil.($t_*>7~10^8$) & $39^{(+0.10)}$&$114^{(+0.74)}$&$331^{(+0.78)}$
&$323^{(-0.21)}$&$500^{(+0.72)}$&$725^{(-0.57)}$ &$1071^{(+0.09)}$ & \\ 
  & & & & 40 &$206^{(+0.21)}$ & $164^{(+0.59)}$ & $21^{(+1.35)}$&{\bf 0.49}\\ 
\hline

$\Gamma_2-1.9$ & $45^{\bf(+1.1)}$&$116^{\bf(+1.0)}$&$322^{(+0.29)}$
&$311^{(-0.87)}$&$491^{(+0.30)}$&$691^{\bf(-1.8)}$&$1061^{(-0.22)}$ & \\
  & & & & 40 &$196^{(-0.66)}$ & $164^{(+0.51)}$ &$23^{(+1.65)}$ &{\bf 0.79}\\

$\Gamma_2-2.0$ & $41^{(+0.52)}$&$125^{\bf(+1.9)}$&$330^{(+0.72)}$
&$331^{(+0.22)}$&$489^{(+0.25)}$&$700^{\bf(-1.5)}$ &$1087^{(+0.58)}$ & \\ 
  & & & & 39 &$205^{(+0.00)}$ & $162^{(+0.43)}$ &$21^{(+1.20)}$ &{\bf 0.64}\\ 

$\Gamma_2-2.2$ & $34^{(-0.58)}$&$115^{(+0.89)}$&$322^{(+0.30)}$
&$319^{(-0.46)}$&$503^{(+0.87)}$&$727^{(-0.48)}$ &$1106^{(+1.2)}$ & \\
  & & & & 45 &$193^{(-0.84)}$ & $166^{(+0.075)}$ &$23^{(+1.8)}$ &{\bf 0.82}\\

$\Gamma_2-2.3$  & $28^{\bf(-1.6)}$&$112^{(0.60)}$&$317^{(+0.01)}$
&$312^{(-0.84)}$&$505^{(+0.97)}$&$764^{(+0.90)}$ &$1124^{\bf(+1.7)}$ & \\
  & & & & 43 &$208^{(+0.24)}$ & $171^{(+1.1)}$ &$21^{(+1.3)}$ &{\bf 0.89}\\ 

$\Gamma_1-1.60$ & $39^{(+0.16)}$&$123^{\bf(+1.7)}$&$327^{(+0.59)}$
&$331^{(+0.24)}$&$507^{(+1.0)}$&$713^{\bf(-1.0)}$ &$1107^{(+1.2)}$ & \\ 
  & & & & 43 &$204^{(-0.04)}$ & $159^{(+0.19)}$ & $22^{(+1.55)}$ &{\bf 0.66}\\ 

$\Gamma_1-1.80$ & $41^{(+0.45)}$&$110^{(+0.43)}$&$315^{(-0.11)}$
&$307^{(-1.1)}$&$479^{(-0.23)}$&$727^{(-0.46)}$ &$1077^{(+0.28)}$ & \\ 
  & & & & 46 &$195^{(-0.73)}$ & $166^{(+0.72)}$ & $23^{(+1.7)}$ &{\bf 0.64}\\ 
\hline

$\gamma 0.03$ & $38^{(+0.03)}$&$113^{(+0.70)}$&$326^{(+0.52)}$
&$316^{(-0.61)}$&$479^{(-0.22)}$&$709^{(-1.1)}$ &$1115^{\bf(+1.4)}$ & \\ 
  & & & & 43 &$206^{(+0.08)}$ &$175^{\bf(+1.4)}$&$20^{(+0.91)}$ &{\bf 0.72}\\ 

$\gamma 0.06$ & $38^{(+0.03)}$&$110^{(+0.37)}$&$318^{(+0.04)}$
&$326^{(-0.08)}$&$478^{(-0.26)}$&$721^{(-0.69)}$ &$1123^{\bf(+1.69)}$ & \\ 
  & & & &46 &$214^{(+0.63)}$&$203^{\bf(+3.6)}$&$25^{\bf(+2.3)}$&{\bf 1.16}\\ 
\hline
\end{tabular}   
\end{table*}

\noindent {\bf -- The synthetic samples:}
Our synthetic stellar samples are computed from first principles: stars 
are created with a random choice of mass $M_*$ and age $t_*$ within a
distribution which is prescribed by the choice of IMF for stellar masses, 
and SFR($t_*$) for the ages. These synthetic stars are then placed on
a fine-meshed grid of individually computed evolution tracks, in order 
to obtain their present-day distribution in the observational 
HR diagram (see also 
Schr\"oder \& Sedlmayr 2001). We computed the evolution tracks with a 
fast but well-tested code, developed by Peter Eggleton 
(Eggleton 1971, 1972, 1973). This code has been updated with 
up-to-date opacities and colour tables. Its performance and 
the right choice of mixing length and overshoot prescription has been 
tested very critically in several studies. Respective evolution models 
have been compared with a number of real MS and giant stars in 
well-observed eclipsing binaries, and isochrones have been compared 
with stellar cluster data (see Schr\"oder et al. 1997, and Pols et 
al. 1997, 1998).

Because of the limited 
observational data, we did not explicitly account for the effects of 
a moderate range in chemical composition, which mostly contribute 
to a more smeared-out distribution of the stars in the synthetic HR
diagram. Instead, we used a random smearing prescription (see Schr\"oder 
1998) which would match the appearance of the observed HR diagram.
It comprises the residual distance uncertainties in the Hipparcos 
parallaxes, photometric errors and some variation in metallicity.

Table 2 lists the characteristic star counts of different
synthetic stellar samples, starting with the best match which we obtained  
with the simple diffusion approximation (model type a), when using
an IMF with $\Gamma_1 = -1.65$ and $\Gamma_2 = -2.05$, an
overshoot-onset (ovo) at $1.50 M_{\odot}$ (see below)
and a constant (genuine) SFR ($\gamma = 0.0$) in the volume.
Obviously, this simplistic 
model gives the best {\it empirical} description of the 
depletion. Therefore, we use it as a reference for all other 
synthetic samples. In all our models (with the exceptions of 
$t_{\rm 9max}10$ and $t_{\rm 9max}8$, see below), 
star formation starts 9 billion years ago, which is a commonly  
adopted age of the thin galactic disk (Hern\'{a}ndez et al, 2001).

\noindent {\bf -- Evaluation of match quality:}
In order to decide the match quality of any synthetic sample on a 
quantitative scale, we compute the average ($|\delta|_{\rm av}$, 
listed by the last column in Table 2) of all individual deviations from 
the observed star counts $n_i$, in units of their individual statistical 
variations $\sqrt{n_i}$. 
We prefer the choice of a simple (linear) average over the rms error 
because not all individual deviations are independent of each other. 
Also, a double weight has 
been given to the deviations of the age-sensitive KGC and LGB counts 
for balancing these with those of the seven MS counts. Of the observed 
complete sample, counts for MS6 and MS7 have been corrected for some 
incompleteness, see previous section. The synthetic star counts were 
obtained from 5 times larger samples, to reduce their statistical 
contributions to the deviations. 

Besides their average, the individual $\delta$-values are equally 
important for the choices of best-matching parameter. In fact,
each of the parameters discussed below affects only certain counts
in a very specific way, while the corresponding change in $|\delta|_{\rm av}$ 
appears to be much less significant. In this way, degeneracies with multiple 
parameter changes are avoided. In Table 2, we therefore 
highlighted individual mismatches by bold-face printing of the 
respective $\delta$-value.

\noindent {\bf -- Onset of overshooting on the MS:}
Because of the now larger (than in previous work) and therefore 
more critical star counts along the MS, we had first to reconsider 
the exact onset of MS overshooting by comparing the effects of 
different evolution grids on the MS star counts.
For this comparison, we used the simple diffusion model because it 
would not introduce any artificial count-change along the MS
(while the artifically abrupt onset of dilution at $t_* >
7 \cdot 10^8$~yrs in our model type (b) does, marginally so). 
The best match (i.e., lowest mean deviation of its star counts from the
observed record) is then achieved with an overshoot-onset (ovo) at 
$1.50 M_{\odot}$, see Table 2. This is in good agreement with 
Pols et al. (1998) but slightly different from our previous choice of
$1.8 M_{\odot}$. 

By contrast, the star counts of synthetic
samples computed with ovo = 1.8, ovo = 1.6, and ovo = 1.4 (all 
parameters for IMF and SFR left unchanged, i.e. $\Gamma_1 = -1.65$,
$\Gamma_2 = -2.05$, $\gamma = 0.0$), show how a different 
onset point on the MS leaves systematic deviations from the observed MS star 
counts just above and below the onset, 
caused by the (then wrong) MS lifetimes of the corresponding stellar models. 
With hindsight, a mismatch in MS4, 
related to the use of an onset at 1.8 $M_{\odot}$, can be noted in
our earlier synthetic samples (Schr\"oder \& Sedlmayr 2001), 
but the much smaller observed $d<50$\,pc sample used then as a reference
did not provide a precise-enough test on the MS overshoot-onset.

\noindent {\bf -- Age of the thin disk population:}
We then made a comparison (again using the simple diffusion model) between
synthetic samples computed for a different age of the ``thin disk''.
These are listed in Table 2 as models $t_{\rm 9max}10$ and  $t_{\rm 9max}8$, 
where star formation starts $10^9$~yrs earlier (see Binney et al.
2000 who argue for an older solar neighbourhood) and  
later. The resulting synthetic samples are not very sensitive to the 
exact value adopted for age of the thin disk. This can be understood, 
since the contribution of the earliest star formation to the 
present-day star counts is depleted by about a factor of 2. 

\noindent {\bf -- Depletion description:}
To demonstrate the importance of matching the depletion factors,
we also list a synthetic sample without any depletion 
($\tau_{\rm dif}=\infty$), which models the solar neighbourhood 
(unrealistically) as a kind of ``closed-box''.
Despite a significant reduction in the IMF of the less massive (older) 
stars to match present-day star counts ($\Gamma_1 = -0.5$ instead of
-1.65), it is impossible to match the observed numbers of 
evolved stars. These are much lower than those predicted by stellar
evolution time-scales -- clearly, a depletion is required which 
progresses with age. 

See Fig. 2 for a plot of the depletion factor as a function of stellar age,
compared to the reciprocal scale-height on an appropriately adapted scale
(1 = 1/195~pc). Both curves coincide well for ages larger than 
$7 \cdot 10^8$yrs, for which the dilution approach applies.

\begin{figure}
\vspace{6.5cm}
\includegraphics{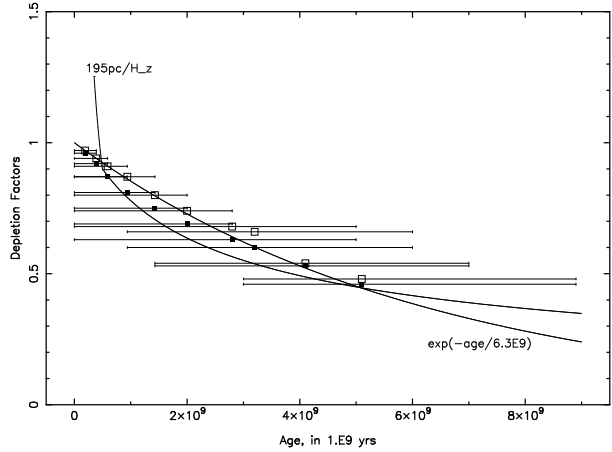}
\caption{Empirical depletion factors, averaged over star groups
         listed in Table 3, with approximate age ranges indicated. 
         Open squares: depletion according to model (a), 
         filled squares: model (b). Shown for reference are the 
         depletion prescriptions for individual age: the exponential
         depletion used by model (a), 
         and the dilution ($\propto H_z^{-1}$) used by model (b) beyond
         the age required for thermalization and radial mixing ($t_* >
         7 \cdot 10^8$yrs).}
\end{figure}

\begin{table*}             
\caption[]{Averaged properties (masses, ages, scale heights, depletion 
factors) with the simple diffusion approximation (a) and with stellar
depletion by (mainly) dilution (b) for the synthetic stars of
specific HRD regions (compare with Table 1).}

\begin{tabular}{l|rrrrrrr|rrr}
\hline 
               & MS1 & MS2 & MS3 & MS4 & MS5 & MS6 & MS7 & KGC & LGB & CW \\
\hline 
$<M>/M_{\odot}$& 4.0 & 2.6 & 2.0 & 1.70& 1.53& 1.37& 1.24& 1.70&1.30&1.53\\   

$<$Age$>$/yrs  & 1.9 $10^8$& 3.9 $10^8$& 5.9 $10^8$& 9.4 $10^8$& 1.43 $10^9$&
                 2.0 $10^9$& 2.8 $10^9$& 3.2 $10^9$& 5.1 $10^9$& 4.1 $10^9$\\

$<H_z>$/pc     &  80 & 135 & 170 & 210 & 250 & 280 & 320 & 350 & 430 & 390 \\ 

$F_{\rm depl}$(a)&0.97&0.94&0.91&0.87&0.80&0.74&0.68&0.66&0.48&0.54\\  

$F_{\rm depl}$(b)&0.96&0.92&0.87&0.81&0.75&0.69&0.63&0.60&0.46&0.53\\
\hline
\end{tabular}   \end{table*}

Finally, Table 2 lists several samples with the above-described approach of
a depletion by dilution. The entry of Dil.($t_*>0.7$~Gyr)  
represents the best match, obtained with $\Gamma_1 = -1.70$ and
$\Gamma_2 = -2.10$ for the IMF and without any change over time of the 
genuine SFR ($\gamma = 0.0$) in the sample volume. Table 3 and Fig. 2 
show that the actual depletion in the various groups of stars of this 
model are indeed very similar to those derived from the simplistic but 
well-matching diffusion 
approximation. The final synthetic samples in Table 2 demonstrate the
sensitivity of the synthetic samples to changes in the parameters for
IMF and SFR. 

\noindent {\bf -- Cut-off for z-values:}
We have also tested our anticipation that it would matter to 
select for stars with small $z$ when deriving
the IMF from a given volume in the galactic plane.
For this purpose, we modelled the stellar content of the {\it whole} 
spherical volume in $d < 100$~pc, i.e., with the larger $z$-values 
included. The best match (not listed here) then indeed requires a 
much steeper IMF in its upper part, with $\Gamma_2 = -2.35$ 
($\Gamma_1 = -1.75$), caused by the already significant drop in 
massive stars with increasing $|z|$. 

\noindent {\bf -- Best matching IMF and SFR:}
Table 4 summarises the properties of the IMF and SFR of the
best matching synthetic samples. Despite their different approach 
to describe the depletion of the galactic plane, both models (a) 
and (b) result in a quite steep IMF with 
$\Gamma_1 = -1.7 \pm 0.1$ for $1.6 M_{\odot} > M_* > \approx 1.1 M_{\odot}$ 
and $\Gamma_2 = -2.1 \pm 0.1$ for $\approx 4 M_{\odot} > M_* > 1.6 M_{\odot}$. 

Furthermore, we find a present star 
formation rate SFR(0) in the solar neighbourhood of about
$3.2 (\pm 0.1) \cdot 10^{-6}$*/yr in a volume of $1.5 \cdot 10^6$pc$^3$ 
centred on the galactic plane, which is 2.1 stars per 1000 years and 
(kpc)$^3$. This value of the SFR applies only to the stars with 
a mass larger than $0.9 M_{\odot}$. We cannot consider any less massive 
stars because these are not luminous enough to appear in any of the 
specific HRD regions counted here, nor would they have evolved into 
the giant branch yet.

In case there may have been a genuinely higher SFR in the past (9 billion 
years ago) on a 1.3 to $1.7 \times$ higher level (averaged radially
and over a time-scale of a billion years), we tried models with $\gamma = 
0.03$ and $\gamma = 0.06$, respectively. Such synthetic samples require a 
less steep IMF (see Table 4), but the over-all quality of the best match
is systematically less good (larger $|\delta_{\rm av}|$) than is achieved 
with the constant level of ``thin disk'' star formation. In particular, 
the count of the oldest group of stars (LGB) becomes significantly too 
large (see last lines in Table 2) when $\gamma$ exceeds 0.03.

Finally, Fig. 3 shows the approximate star formation rate in the 
column (acc. to Equ. 15), as it appears from our observations in 
the volume in combination with our semi-empirical density scale-height
as a function of age (Equ. 10, 11). As already mentioned,
the column integral SFR$_{\rm col}$ grows by a factor of 4 from its
low present-day (local, inbetween spiral arms) value until it reaches an 
approximately constant, non-local disk-average of 0.82~*~Gyr$^{-1}$pc$^{-2}$
for all stars older than about $7 \cdot 10^8$yrs (and $M_* > 0.9 M_{\odot}$). 
Of these, a large fraction must have been collected by the present-day 
solar neighbourhood as a result of radial mixing. We would like to note 
that a sampling range on the kpc scale (like the one for luminous stars 
of Scalo, 1986) could not show such a pronounced effect. Rather, a 
large sample volume partly includes the neighbouring spiral arms and 
should show SFR$_{\rm col}$ values much nearer to the disk-average
in the first place.

\begin{figure}
\vspace{6.5cm}
\includegraphics{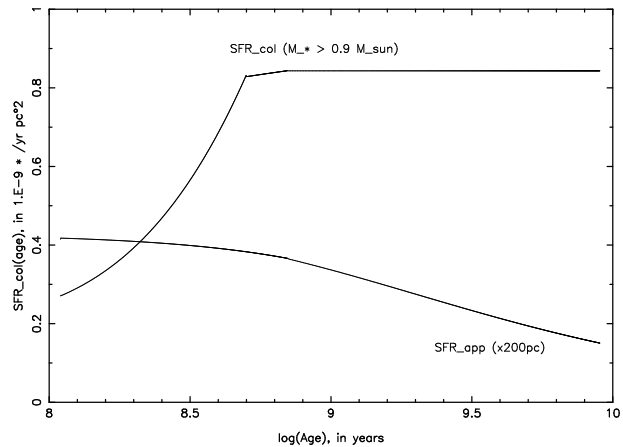}
\caption{Column-integrated SFR$_{\rm col}$ of the galactic disk
       over stellar age derived from our model (b), given in
       $10^{-9}$ stars per year and pc$^2$, over log(age).
       Note the apparent increase over the recent past 
       ($t_* < 7 \cdot 10^8$yrs) where radial mixing changes 
       the observed SFR from its low, local (interarm) value 
       to a much larger (and fairly constant) non-local average. 
       For reference, the apparent SFR in the local volume is also shown 
       ($\times 200$~pc). Its decline with age reflects the dilution of
       ``thin disk'' stars as they expand into larger scale-heights.} 
\end{figure}

\begin{table}             
\caption[]{Slope of IMF (Scalo notation) listed for the two best matching 
models (simple diffusion approximation and depletion by diffusion \& 
dilution, see Table 2), both with constant SFR, and two less good 
matches using a SFR slightly higher in the past.}

\begin{tabular}{lrrrrr}
\hline 
Depletion by & SFR(0) & $\gamma$  
          & $\Gamma_1$ & $\Gamma_2$  & $|\delta|_{\rm av}$ \\
\hline 
$\tau_{\rm dif}=6.3~10^9$yrs   & $3.03~10^{-6}$ & 0.00 & 
                     -1.65 & -2.05 & {\bf 0.30} \\
Dil.($t_*>7~10^8$)& $3.2~10^{-6}$&0.00&-1.70&-2.10&{\bf 0.49}\\
\hline
Dil.($t_*>7~10^8$)& $3.0~10^{-6}$&0.03&-1.60&-2.00&{\bf 0.72}\\
Dil.($t_*>7~10^8$)& $2.8~10^{-6}$&0.06&-1.40&-2.00&{\bf 1.16}\\
\hline
\end{tabular}   \end{table}


\section{Discussion and outlook}

In our approach to assess the local galactic ``thin disk'' IMF of
single field stars by means of synthetic stellar samples which
match the observed, complete (volume-limited) sample of solar 
neighbourhood stars, 
we have given special attention to three points vital to
an unbiased IMF assessment: discrimination for stars near the
galactic plane ($|z| < 25$~pc), a realistic (semi-empirical) 
quantification of the stellar depletion in the galactic plane,
and use of the specific, age-dependent information buried in 
the account of the evolved stars in the stellar sample.

Although we demonstrated that the simple approach of an age-independent
diffusion time-scale, as used by us before (Schr\"oder \& Sedlmayr, 2001), 
does not yield too much different an IMF from the more sophisticated, 
semi-empirical depletion by diffusion and dilution, the IMF derived 
here supersedes our earlier work in several respects:
(i) The present work is based on a much larger observed sample and 
yields significantly smaller error bars. (ii) The observed sample is
a better representation of the stellar component in the immediate galactic 
plane, and (iii) the synthetic samples have now been created with a better 
choice of the onset of overshooting on the MS.

The slope of the field star IMF (i.e., $\Gamma_1 = -1.7$, $\Gamma_2 = -2.0$) 
obtained by this study for medium to high masses is remarkably similar 
to the range of values considered by Scalo (1986, 1998), despite being 
based on a very different approach. 
In principle, the genuine IMF and the depletion-corrected SFR 
in a defined volume on the galactic plane, as determined here, 
are equivalent to their column-integrated quantities studied
by the classical approach, in how they depend on mass and age. 
In fact, in our column-integrated quantities, a depletion by simple
geometrical dilution must cancel with the growing scale height.
At least, this is true for stars older than $7 \cdot 10^8$yrs. 

However, for stars younger than that, our column-integrated values are 
changing considerably, as demonstrated for the SFR by Fig. 3. This is
the result of a progressive radial mixing into the small, local
and inter-spiral-arm volume used here. For the respective stars (mainly,
$M_* > 2 M_{\odot}$) our column-integrated IMF would be unphysically
steep. It does not compare to Scalo's (1986) IMF because his sampling 
range is about an order of magnitude larger. Also, he uses different
scale heights for that range of young stars. Still, upon closer inspection
(Pagel, 1997, p.~207), a slightly steeper slope ($\Gamma = -2.4$) does 
actually appear in the respective data of Scalo (1986). This may be due 
to the same effect (radial mixing -- hence, not genuine), 
but much reduced by his vastly larger sampling scale. 
In any case, the slope of the lower mass IMF remains unaffected.

We see our approach of monte-carlo-created stellar samples, based on
real evolution tracks and a good semi-empirical quantification of stellar 
depletion, compared one-to-one with an observed complete sample, as 
complementary to recently developed methods of inverting colour-magnitude
diagrams with a maximum entropy method (see Hern\'{a}ndez et al. 1999).
While both methods make use of the {\it full} information content of the 
HR diagram, we find our approach is a very versatile tool and it may 
be more robust when studying the impact of more than one parameter. 

Still, the above-mentioned problems with radial mixing and a very limited 
sampling-range make it
difficult to deduce an exact slope of the galactic disk IMF for the
more massive stars. In fact, we are unable to adress the important question
whether it may turn into a Salpeter dependence for the most massive
stars (beyond about $6 M_{\odot}$) which are important
for the UV radiation field and any interaction with the ISM.
Rather, the size of our observed sample is  
more limited than we would wish for. This is also true when it comes to
the question of any genuine change of the SFR, or even the IMF, over  
time (see discussion by Kroupa 2001, e.g. ). The results presented in the 
previous section provide, at best, support for an approximately constant 
SFR. 

But these limitations will disappear as the situation is going to 
change dramatically in about a decade, when a new dimension of data, 
in a comfortably large sampling range, will
become available with the above-mentioned projects GAIA and DIVA. 
Well tested, detailed models for creating synthetic stellar 
populations will then be a very powerful means of assessing the 
star formation history in our galaxy, and the depletion by dilution
which provides a sensitive test of our understanding of galactic 
dynamics. 





\begin{thebibliography}{}
\bibitem{} Bastian U., R\"oser S., Hoeg E, Mandel H., et al. 1996, 
             Astron. Nachr. 317, 281
\bibitem{} Bertelli G., Mateo M., Chiosi C., Bressan A., 1992, ApJ 388, 400
\bibitem{} Bertelli G., Nasi E. 2001, ApJ 121, 1013
\bibitem{} Binney J., Dehnen W., Bertelli G. 2000, MNRAS 318, 658
\bibitem{} Burkert A., Truran J.W., Hensler G. 1992, ApJ 391, 651
\bibitem{} Cohen M. 1995, ApJ 444, 874
\bibitem{} Conti P.S., Vacca W.D. 1990, AJ 100, 431
\bibitem{} Dwek E. 1998, ApJ 501, 643 
\bibitem{} Eggen O.J., Lynden-Bell D., Sandage A.R. 1962, ApJ 136, 748
\bibitem{} Eggleton P.P., 1971, MNRAS 151, 351
\bibitem{} Eggleton P.P., 1972, MNRAS 156, 361
\bibitem{} Eggleton P.P., 1973, MNRAS 163, 179
\bibitem{} Gilmore G., Reid I.N. 1983, MNRAS 202, 1025
\bibitem{} Gilmore G. 1999, Ap\&SS 267, 109
\bibitem{} Hern\'{a}ndez X., Valls-Gabaud D., Gilmore G. 1999, MNRAS 304, 705
\bibitem{} Hern\'{a}ndez X., Avila-Reese V., Firmani C. 2001, MNRAS 327, 329
\bibitem{} Hensler G. 1999, Ap\&SS 265, 397
\bibitem{} Jeans J.H. 1915, MNRAS 76, 71
\bibitem{} Jura M., Kleinmann S.G. 1992a, ApJS 79, 105
\bibitem{} Jura M., Kleinmann S.G. 1992b, ApJS 83, 329
\bibitem{} Jura M. 1994, ApJ 422, 102
\bibitem{} Kent S.M., Dame T.M., Fazio G. 1991, ApJ 378
\bibitem{} Kerber L.O., Javiel S.C., Santiago B.X. 2001, A\&A 365, 424
\bibitem{} Kroupa P., Tout C.A., Gilmore G., 1991, MNRAS, 251, 293
\bibitem{} Kroupa P. 2001, MNRAS 322, 231
\bibitem{} Kuijken K., Gilmore G. 1989, MN 239, 605  
\bibitem{} Lacey C. 1984, MNRAS 208, 687
\bibitem{} Layden, Andrew C. 1995, AJ 110, 2288
\bibitem{} Lynden-Bell D. 1962, MNRAS 124, 1
\bibitem{} Marsakov V.A., Shevelev Y.G. 1995, AZ 72, 630
\bibitem{} Mendez R.A., van Altena W.F. 1996, AJ 112, 665
\bibitem{} Miller G.E., Scalo J.M. 1979, ApJS 41, 513
\bibitem{} Oestreicher M.O., Schmidt-Kaler T. 1995, A\&A 294, 57
\bibitem{} Ojha D.K., Bienaym\'{e} O., Robin A.C., Cr\'{e}ze M., 
             Mohan V. 1996, A\&A 311, 456
\bibitem{} Ojha D.K., Bienaym\'{e} O., Mohan V., Robin A.C. 1999, 
             A\&A 351, 945
\bibitem{} Oort, J.H. 1932, Bull. Astr. Inst. Netherlands 6, 249
\bibitem{} Pagel B.E.J. 1997, in ``Nucleosynthesis and Chemical Evolution
             of Galaxies'', CUP, Cambridge
\bibitem{} Pagel B.E.J. 2001a, in ``Cosmic Evolution'', Vangioni-Flam E.,
             Ferlet R., Lemoine M (eds.), New Jersey: World Scientific, 223
\bibitem{} Pagel B.E.J. 2001b, PASP 113, 137
\bibitem{} Perryman M.A.C., Lindegren L., Kovalevsky J., H{\o}g E., Bastian U.,
           et al. 1997, A\&A 323, L49
\bibitem{} Perryman M.A.C. 2002, Ap\&SS 280, 1 
\bibitem{} Phleps S., Meisenheimer K., Fuchs B., Wolf C. 2000, A\&A 356, 108
\bibitem{} Pilyugin L.S. 1996, A\&A 313, 803
\bibitem{} Pols O.R., Tout C.A., Schr\"oder K.-P., Eggleton P.P., Manners J. 
             1997, MNRAS, 289, 869
\bibitem{} Pols O.R., Schr\"oder K.-P., Hurley J.R., Tout C.A., Eggleton P.P. 
             1998, MNRAS, 298, 525
\bibitem{} Reed B.C., 2000, AJ 120, 314
\bibitem{} Robin A.C., Haywood M., Cr\'{e}ze M., Ojha D.K., Bienaym\'{e} O. 
             1996, A\&A 305, 125
\bibitem{} Scalo J.M. 1986, Fundam. Cosmic Phys. 11, 1
\bibitem{} Scalo J.M. 1998, in ``The Stellar Initial Mass Function'', ASP Conf.
             Series 142, Gilmore G., Howell D. (eds.), 201
\bibitem{} Schr\"oder K.-P., Pols O.R., Eggleton P.P. 1997, MNRAS 285, 696
\bibitem{} Schr\"oder K.-P. 1998, A\&A 334, 901
\bibitem{} Schr\"oder K.-P., Sedlmayr E. 2001, A\&A 366, 913
\bibitem{} Seifert W., Mandel H., Wagner S., Bastian U., Roeser S. 1998,
              SPIE 3356, 904
\bibitem{} Siebert A., Bienaym\'e O., Soubiran C. 2002, A\&A, in print
\bibitem{} Sommer-Larsen J., Antonuccio-Delugo V. 1993, MNRAS 262, 350
\bibitem{} Vergely J.-L., Ferrero R.F., Egret D., K\"oppen J. 1998, 
             A\&A 340, 543
\bibitem{} Wielen R. 1977, A\&A 60, 263 
\end{thebibliography}
\end{document}